\documentclass[twocolumn]{emulateapj}
\usepackage{color, natbib}
\shorttitle{Rigel Variability}
\shortauthors{Moravveji et al.}

\newcommand{\most}{\textit{MOST}}

\newcommand{\teff}{T_{\mbox{\scriptsize eff}}}
\newcommand{\dhip}{d_{\mbox{\scriptsize Hip}}}

\newcommand{\kap}{$\kappa$-mechanism}

\begin{document}
\title{Asteroseismology of the Nearby SN-II Progenitor: Rigel \\
Part \textsc{I}. The \most\footnote{Based on data from the \most ~satellite, a Canadian Space Agency mission,
operated jointly by Dynacon, Inc., the University of Toronto Institute of Aerospace
Studies, and the University of British Columbia, with the assistance of the
University of Vienna.}  ~High Precision Photometry and Radial Velocity Monitoring} 


\author{Ehsan Moravveji\altaffilmark{1,2}} 
\altaffiltext{1}{Department of Physics, Institute for Advanced Studies in Basic Sciences (IASBS), Zanjan 45137-66731, Iran}
\email{moravveji@iasbs.ac.ir}
\author{Edward F. Guinan\altaffilmark{2}}
\altaffiltext{2}{Department of Astronomy, Villanova University, 800 Lancaster Ave, Villanova, PA 19085, USA}

\author{Matt Shultz\altaffilmark{3}}
\altaffiltext{3}{Royal Military College of Canada, PO Box 17000, Station Forces, Kingston, ON K7K 4B4, Canada}

\author{Michael H. Williamson\altaffilmark{4}}
\altaffiltext{4}{Center of Excellence in Information Systems, Tennessee State University, Nashville, USA}

\author{Andres Moya\altaffilmark{5}}
\altaffiltext{5}{Departamento de Astrof\'{\i}sica, Centro de Astrobiolog\'ia (INTA-CSIC), PO BOX 78, 28691 
Villanueva de la Ca\~nada, Madrid, Spain}
\date{last updated: \today}
\begin{abstract}
Rigel ($\beta$ Ori, B8 Ia) is a nearby blue supergiant displaying $\alpha$ Cyg type variability, and is one of the nearest type-II supernova progenitors.
As such it is an excellent test bed to study the internal structure of pre core-collapse stars. 
In this study, for the first time, we present 28 days of high precision \most ~photometry and over 6 years of spectroscopic monitoring. 
We report nineteen significant pulsation modes of SNR$\gtrsim$4.6 from radial velocities, with variability time 
scales ranging from 1.21 to 74.7 days, which are associated with high order low degree gravity modes. 
While the radial velocity variations show a degree of correlation with the flux changes, there is no clear interplay between the equivalent widths of 
different metallic and H$\alpha$ lines. 
\end{abstract}
\section{Introduction}\label{s:intro}

Blue supergiants (BSGs) such as Rigel are post-main sequence (MS) massive stars ($M\gtrsim 15M_\odot$) evolving quickly 
across the Hertzsprung-Russel diagram. 
They are amongst the most intensively studied objects in contemporary astrophysics \citep[see][for a recent review]{maeder-2011-01-arXiv}.
BSG stars are intriguing objects because they end their lives as Type II supernovae (SN-II) which serve as catalysts for star formation and 
primary contributors to the heating and chemical enrichment of their host galaxies. 
These stars also offer a potentially valuable addition to the distance ladder by way of the BSG wind-momentum-luminosity \citep{kudritzki-1999-01} and 
flux-weighted gravity-luminosity relationships \citep{kudritzki-2003-01} which give very consistent distance determinations to the Local Group of galaxies 
\citep{kudritzki-2008-01,vivian-2009-01}.  

Fascinating features of BSGs include their microvariability in flux \citep[e.g.][]{sterken-1977-01}, complex changes in radial velocity (RV),
as well as variations in the equivalent widths (EW) and line profiles of Balmer (especially H$\alpha$) and metallic lines \citep{kaufer-1996-01, kaufer-1997-01,
richardson-2011-01}. 
Kaufer and coworkers monitored the optical spectra of six BA supergiants (including Rigel and $\alpha$ Cyg). 
They analyzed the radial velocity measures of these stars and associated their observed variability with non-radial pulsations since the traveling features in the 
dynamical spectra could not be reconciled with the rotational periods of the stars. 
\citet{waelkens-1998-01} discovered 32 pulsating BSGs in the Hipparcos database which belong to the $\alpha$ Cyg class of variable stars, with 
periods ranging from 1.5 to 24 days. 
They were later definitely identified as gravity mode oscillations by a Non-LTE spectroscopic analysis of \citet{lefever-2007-01} who compared the 
current position of their stars with the instability strip of g-mode dominated pulsators. 
Beyond the Milky Way, \citet{bresolin-2004-01} detected a handful of variable BSGs in NGC 300, and found the periods of two of these stars to be 72.5 and 96 days. 
Recently, \citet{aerts-2010-02} correlated the sudden amplitude decline in the spectroscopically peculiar CoRoT target HD 50064 (B1-6 Ia) to radial 
\textit{strange mode} 
variability with a 37 day period; they inferred a mass of $\sim45M_\odot$ for this luminous mass-losing BSG.
Despite several decades of observational efforts on ground based photometry and spectroscopy of $\alpha$ Cyg type pulsating stars, 
it is not certain what portion of their variability is periodic, nor how far they deviate from strict periodicity.

The nature of the aforementioned variability in high mass post-MS variables is still poorly understood. 
The photometric variability of HD 163899 (B2 Ib/II) found by \most ~observations revealed a total of 48 pressure (p) and gravity (g) modes. 
\citet{saio-2006-01} showed that these modes can be simultaneously excited by the \kap ~and reach the surface if they arrive at the base of the 
intermediate convection zone (ICZ) with an appropriate phase. \citet{gautschy-2009} searched for the origin of long-period variabilities in the 
prototype $\alpha$ Cyg. 
Interestingly his Figure 5 shows a gap where no instabilities are predicted for evolutionary tracks with $3.95\lesssim\log\teff\lesssim4.15$. 
Rigel lies in this gap. 
In contrast to this result, the study of \citet{saio-2011-01} predicts that Rigel should be unstable against non-radial convective g$^-$ modes.
This contradiction could arise from different physical ingredients (such as rotational and overshoot mixing) and various numerical techniques
in pulsation codes.
This is not surprising since evolved massive stars like Rigel are demanding to model. 

The radial strange mode is proposed as another mechanism to induce instability and interplay with mass loss efficiencies in these 
massive stars with $\log(L/M)\gtrsim4\log(L_\odot/M_\odot)$ \citep[see e.g.][and references therein]{dziembowski-2005-01,saio-2011-01, aerts-2010-02}. 
This requirement is also marginally fulfilled by Rigel. \citet{godart-2009-01} investigated the destructive impact of core overshooting and mass loss 
during the MS lifetime on the extent of the ICZ, and showed that models with wider ICZs are more likely to destabilize stellar oscillations. 
These studies show that asteroseismology of slowly pulsating blue supergiants can reveal a wealth of information about the internal structure of SN-II 
progenitors. 
This can be complementary to the understanding of the physical properties of pulsars which originate from the cores of massive stars \citep{heger-2005-01}.

This is the first paper in a series of investigations aiming at probing the details of the internal structure of BSGs through asteroseismic study,
and investigating the origin of their short- and long-period variability. 
Rigel was selected as an ideal test subject because of its apparent brightness, proximity and current evolutionary status. 
The latest measurements of the physical parameters of Rigel collected from the literature are summarized in Section \ref{s:rigel}. 
The space-based \most ~photometry and ground-based spectroscopy are presented in Section \ref{s:obs}. 
The results of the multimode pulsation frequencies are finally discussed in Section \ref{s:freq}. 
The interpretation of the pulsation frequencies will appear in Moravveji et al. (in preparation). 
%
\begin{figure*}[t!]
\epsscale{1.2} \plotone{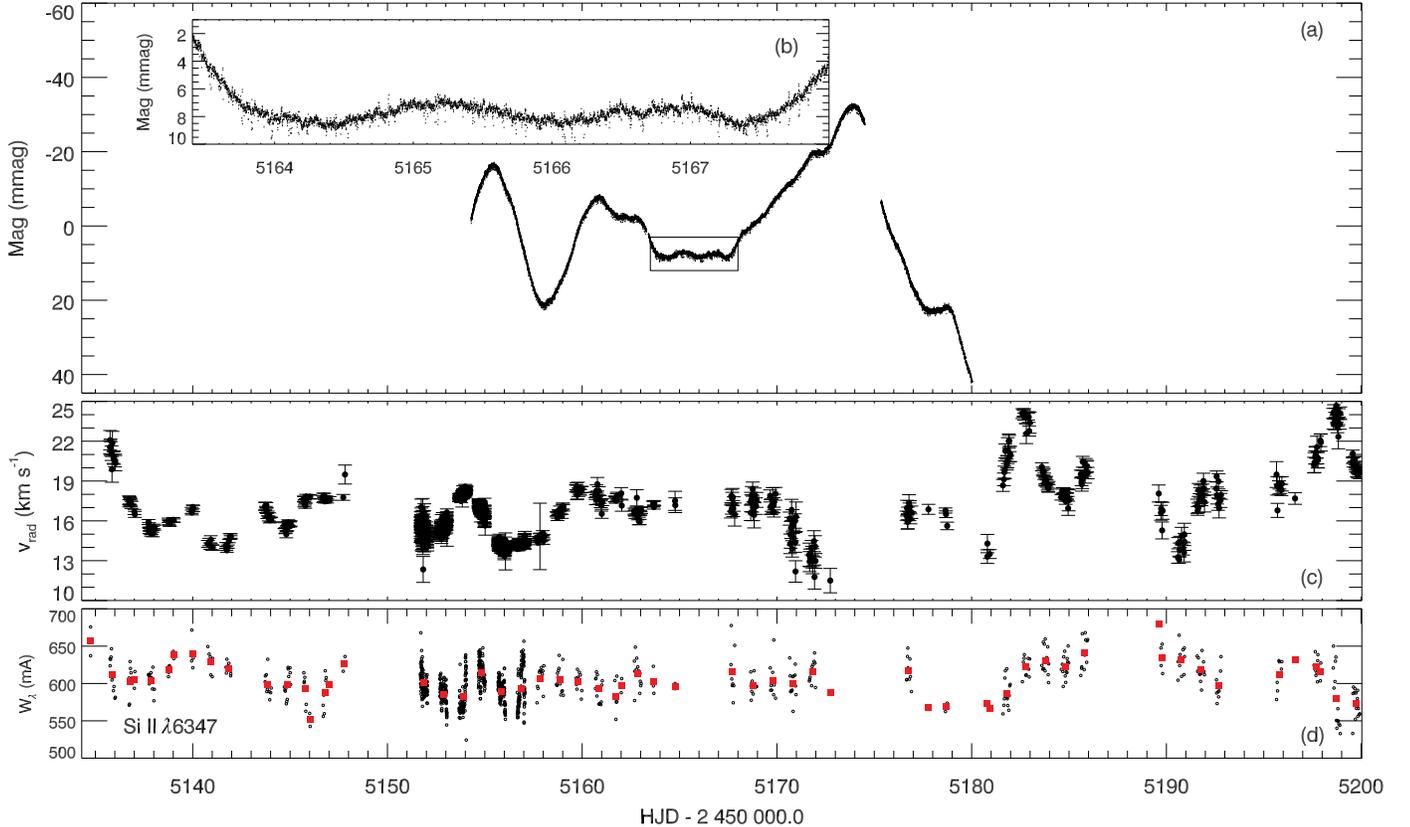}
\caption{Compilation of observations from the Rigel campaign.
Panel (a) shows the 27.7-day \most ~space photometry (Section \ref{ss:most}).
Panel (b) is a  5-day zoomed-in portion of the data to emphasize the amplitude and time scale of the shortest variations.
Panel (c) shows the RV data which demonstrates a correlation with the flux changes (Section \ref{ss:spec}).
Panel (d) is the equivalent width variation for Si II $\lambda$6347.1 $\AA$ during the same monitoring epoch (Section \ref{ss:ew}); 
the (red) squares are the 1-day binned averages (see Figure \ref{f:residuals}).
}\label{f:rv-lc}
\end{figure*}
\section{About Rigel}\label{s:rigel}
Rigel ($\beta$ Ori; HD 34085; B8 Ia; V$\sim$0.12 mag) 
is the 6$^{\mbox{\scriptsize th}}$ brightest star in the night sky and the most luminous star in the solar neighborhood.  
It is a member of a multiple star system where its companion, Rigel B, is a spectroscopic binary about 9.5\arcsec distant \citep{sanford-1942-01}.
From the revised Hipparcos parallax the distance to Rigel is
$\dhip=264\pm24$ pc \citep{vanleeuwen-2007-01} which is smaller than $360\pm40$ pc adopted by \citet{hoffleit-1982-01} 
who assumed that Rigel originates from the $\tau$ Ori R1 complex.

Fundamental parameters of Rigel which impose valuable constraints on the equilibrium model of the star are already measured. 
They are effective temperature $T_{\mbox{\scriptsize eff}}=12\,100\pm150$ K, surface gravity $\log g=1.75\pm0.10$, 
luminosity $\log (L/L_\odot)=5.08^{+0.07}_{-0.10}$, near-solar metalicity [M/H]=$-0.06\pm0.10$, 
surface He abundance $Y_{\mbox{\scriptsize s}}=0.32\pm0.04$, $v\sin i\approx25\pm3$ km s$^{-1}$ 
\citep{przybilla-2006-01, przybilla-2010-01, 
simon-diaz-2010-01}. The most up-to-date limb darkened angular diameter for Rigel comes from CHARA/FLOUR K-band 
interferometry  $\theta_{\mbox{\scriptsize LD}}=2.75\pm0.01$ mas \citep{aufdenberg-2008-01} which in combination with $\dhip$ 
yields $R=78.9\pm7.4 R_\odot$.
\citet{przybilla-2010-01} propose a mass of 23$M_\odot$ from Geneva evolutionary tracks which include the effects of rotation.

The absolute bolometric magnitude is calculated (adopting the observed $B-V=-0.03$ from \citet{nicolet-1978-01}, 
and a bolometric correction $BC=-0.78$ from \citet{bessell-1998-01} with $E(B-V)=0.05$ from \citet{przybilla-2006-01}) to yield 
$M_{\mbox{\scriptsize bol}}=-7.92\pm0.28$.
By applying the flux-weighted gravity-luminosity relation of \citet{kudritzki-2003-01} 
$$M_{\mbox{\scriptsize bol}}=3.71\, \log (g/T_{\mbox{\scriptsize eff, 4}}^4) - 13.49$$
($T_{\mbox{\scriptsize eff, 4}} = \teff /10\,000$) which is optimized for BSGs,
we arrive at a value of $M_{\mbox{\scriptsize bol}}=-8.17\pm0.45$ - in good agreement with our calculated value.

Moreover, Rigel's line profile variability in H$\alpha$ \citep{kaufer-1996-01, morrison-2008-01} and other metal-line equivalent widths due to 
non-radial pulsations \citep{kaufer-1997-01} and mass loss \citep{chesneau-2010-01} is already published.

\par Rigel shows variations in the H$\alpha$ spectral feature with some outburst events being recorded by \citet{israelian-1997-01} 
and Chesneau (VLTI campaign, in preparation). 
The morphology of this line is studied by \citet{morrison-2008-01}. 
Moreover, \cite{chesneau-2010-01} detected an extended rotating H$\alpha$ region 
from optical interferometry with VEGA/CHARA \citep{mourard-2009-01} and speculate that Rigel is observed with 
its rotation axis along the North-South direction at significantly high inclination angle.
They estimated the current mass loss rate is $\dot{M}=1-2\times10^{-7} M_\odot$ yr$^{-1}$.
Assuming a high inclination angle, the rotation velocity is well below 20\% of the estimated critical breakup rotation rate of $\sim$185 km s$^{-1}$
\citep[Eq. 2.19 in][]{maeder-2009-book}, so spherical symmetry can be safely assumed for its geometry.
The current best values of the physical parameters of Rigel are presented in Table \ref{t:data}. 
Additional parameters for Rigel are given by \citet{przybilla-2006-01} .
These are invaluable input for future seismic modeling and analysis.
\begin{deluxetable}{lcl} 
\tablecolumns{3}
\tabletypesize{\normalsize} 
\tablecaption{Updates to Physical Parameters of Rigel \label{t:data}} 
\tablehead{ \colhead{Parameter} & \colhead{Value} & \colhead{Reference} }
\startdata 
$\dhip$ [pc]  & $264\pm24$  & \citet{vanleeuwen-2007-01} \\
$\teff$    [K]   & $12\,100\pm150$  & \citet{przybilla-2010-01} \\
$\log(L/L_\odot)$  & $5.08^{+0.07}_{-0.10}$  & This work \\
$Y_s$ & $0.32\pm0.04$  & \citet{przybilla-2010-01} \\
$v\sin i$  [km s$^{-1}$] & $25\pm3$  & \citet{przybilla-2010-01} \\
  &  & \citet{simon-diaz-2010-01} \\
Inclination  & $60^\circ \lesssim i \lesssim 90^\circ$  & This work \\
$\theta_{\mbox{\scriptsize LD}}$ [mas]  & $2.75\pm0.01$ & \citet{aufdenberg-2008-01} \\
$R/R_\odot$  & $78.9\pm7.4$ & This work \\
$M_{\mbox{\scriptsize bol}}$ [mag]  & $-7.92\pm0.28$  & This work \\
$\dot{M}$  [$M_\odot$ yr$^{-1}$] & 1-2$\times10^{-7}$  & \citet{chesneau-2010-01} \\
$B_d$ [G] & $\lesssim$ 25  & \citet{shultz-2011-01} \\
$P_{\mbox{\scriptsize shortest}}$ [d]  & $1.2191\pm0.0001$  & This work \\
$P_{\mbox{\scriptsize longest}}$  [d] & $74.74\pm0.28$            & This work \\
\enddata 
\end{deluxetable} 
%
\section{Observations}\label{s:obs}
\subsection{\most ~High Precision Photometry}\label{ss:most}
Rigel was observed continously with the \most ~satellite \citep[Microvariability and Oscillations of Stars,][]{walker-2003-01} for 27.7 days 
from 15 November to 13 December 2009. 
The dataset consists of 30\,640 observations after correcting for the Southern Atlantic Anomaly.
Figure \ref{f:rv-lc}.a shows the light curve after removing an offset of $\sim0.12$ mag.
The abscissa is in HJD$-$2\,450\,000.0. 
Because of the brightness of the target, the precision of the observations is of the order of $\approx$0.05 to 0.10 mmag.
Starting with a sinusoidal variability pattern,
oscillations die out and Rigel seems quiescent for several days, approximately from HJD 5164 to 5168 (enlarged in Figure \ref{f:rv-lc}.b), 
followed by a gradual rise, and a steep decline in light flux; 
this resembles a possible beating pattern and we expect close, low-frequency modes to appear in the harmonic analysis (Section \ref{s:freq}).
There is a 20-hour gap around HJD 5175, arising from an interruption in communication between
the satellite and the ground station. This is at the time when the star is decreasing in brightness almost monotonically. 
%
\begin{figure}[th!]
\epsscale{1.2} \plotone{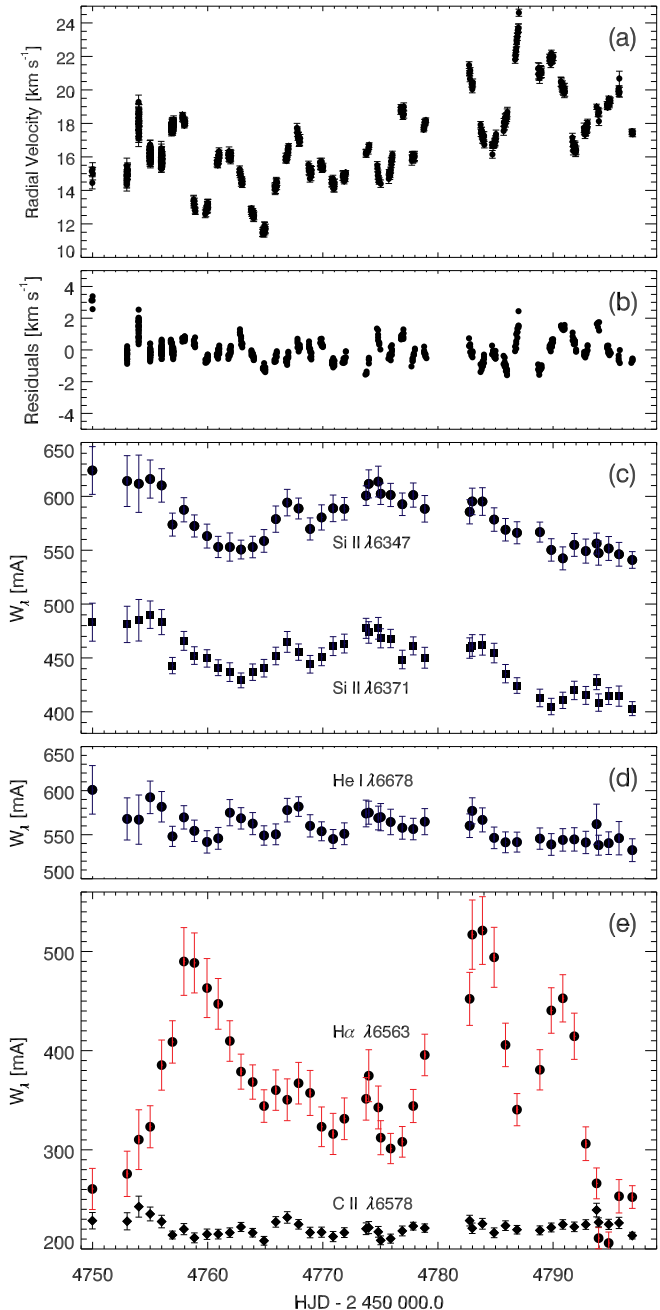}
\caption{(a) A representative  50 day radial velocity curve of Rigel. 
1-$\sigma$ error in RV is overplotted. 
(b) A residual plot with 19 harmonics removed as in Table \ref{t:freq} with rms deviation 0.59 km s$^{-1}$.
(c) The 1-day binned EW measures of two metallic lines, namely Si \textsc{II} $\lambda\, 6347$ and Si \textsc{II} $\lambda\, 6371\, \AA$
which exhibit significant temporal variations.
(d) A similar plot for the He \textsc{I} $\lambda\,6678$ line, and (e) for C \textsc{II} $\lambda\,6578$ and H$\alpha$ lines.
\label{f:residuals}} 
\end{figure}
%
\subsection{Spectroscopy}\label{ss:spec}
Rigel's optical spectrum has been monitored by the 2-m Tennessee State University Automatic Spectroscopic
Telescope at Fairborn Observatory, Arizona \citep[TSU:AST,][]{eaton-2007-01}.
A total of 2\,328 high-resolution (R$\sim$30\,000 and 20\,000) moderate signal-to-noise (SNR$\sim$50 to 150) echelle spectra 
(4\,900 - 7\,100 $\AA$) were obtained.
The spectra were secured  over 334 nights starting from 11 December 2003 to 14 February 2010. 
The density of time-sampling was increased during the months centered on the \most ~observations.
Simultaneous with \most, 442 spectra were collected over 20 nights.

The spectra were reduced at TSU by an automatic pipeline method without removing the telluric lines.
The radial velocity variations were derived by fitting a Gaussian to 29 metallic lines; 
Figure \ref{f:rv-lc}.c shows the RV variations contemporaneous with \most ~observations.
An offset of $+0.3$ km s$^{-1}$ was applied to the AST instrumental RV measurements \citep{eaton-2007-01}.
The standard deviation in the mean for each RV measurement was evaluated as the standard deviation of the 29 
measured absorption lines in each spectrum. 
Averaged over all spectra, this value is 0.26 km s$^{-1}$.
In the years prior to the \most ~observations, Rigel was observed sporadically with AST.
Also, the AST dataset is interrupted by cloudy weather that occurred during the MOST observing run. 
Thus the sampling is not even, and the analysis of these data will suffer from daily and annual aliasing.

Extensive and highly variable velocity changes are clearly visible in Figure \ref{f:rv-lc}c.
During the plotted interval, the RVs vary from 11 to 25 km s$^{-1}$. 
There is evidence for a correlation between the RV and \most ~brightness variations. 
During this interval, and other seasons, the RV variations appear to exhibit complex periodicity typical of a multi-periodic  pulsating star.

In addition to the AST spectroscopy,  Rigel was observed with the VLTI during the time of the \most ~observations. 
A preliminary analysis of the spectro-interferometry by Chesneau et al. (in preparation) shows that 
during this interval Rigel was relatively quiet, 
with the H$\alpha$ line appearing in absorption with sometimes a weak emission component appearing at times, and no outbursts being recorded.
No major high velocity absorption (HVA) event is identified during this interval.
This is confirmed by seven raw Ritter echelle spectra (R = 26,000, SNR 100 to 200) taken during the same interval at 
Ritter Observatory (Morrison; private communications).
As a result, there is no evidence of a possible propagating atmospheric shock or chaotic activity during \most ~photometry.  

\subsection{Spectropolarimetry}\label{ss:spec-pol}
In conjunction with MOST photometry, 78 observations were collected with the high-resolution (R$\sim$65\,000) 
spectropolarimeter ESPaDOnS, installed at CFHT, and its clone Narval at Telescope Bernard Lyot (TBL, Pic du Midi). 
Analysis with the multi-line cross-correlation technique Least Squares Deconvolution \citep[LSD;][]{donati-1997-01} reveals no 
evidence of a magnetic field \citep{shultz-2011-01}, with error bars on the longitudinal field $B_l$ of order 15 G. 
Matching synthetic disk-integrated Stokes V profiles to the observed Stokes V profiles has constrained the dipolar magnetic field to be 
$B_d\lesssim$ 25 G for for high inclinations of the rotational axis and low obliquities of the magnetic axis, although $B_d\sim$ 50 G 
is possible at low-intermediate inclinations, in which case the rotational period is shorter and fewer observations can be binned together 
(Shultz et al., in preparation). 
If, as discussed above, Rigel is viewed nearly equator on, then the first of the conditions necessary for a higher upper limit applies. 
However, these upper limits cannot rule out the possibility of significant magnetic effects. 
Thus, in the absence of any positive evidence for a magnetic field, or any pressing theoretical reason to suspect its existence, 
there is no reason at this time to complicate the pulsational analysis with its inclusion.

\subsection{Equivalent Widths}\label{ss:ew}
\citet{kaufer-1997-01} showed that different spectroscopic lines in BA supergiants emerge from different optical depths $\tau_\lambda$,
in a sense that lines from shallower photospheric depths have larger equivalent widths (hereafter EW designated by $W_\lambda$). 
They classified the lines in the optical band accordingly to weak $(50 \lesssim W_\lambda \lesssim 200$ m$\AA)$, medium 
$(200 \lesssim W_\lambda \lesssim 500$ m$\AA)$ and strong $(W_\lambda \gtrsim 500$ m$\AA)$.  
We chose C \textsc{II} $\lambda6583$ as  a weak line,  H$\alpha$, 
C \textsc{II} $\lambda6578$ and Si \textsc{II} $\lambda6371$ as medium lines,
and He \textsc{I} $\lambda6678$ and Si \textsc{II} $\lambda6347$ as strong lines.

EWs of the above lines are calculated according to 
\begin{equation}
W_\lambda = \int (1-F_\lambda)\, d\lambda.
\end{equation}
\noindent where $F_\lambda$ is the flux at each wavelength interval d$\lambda$, renormalized to the continuum on either side of the spectral line. 
To suppress the contribution from cosmic rays the spectra were filtered using a low-bandpass median filter; 
contamination from telluric lines was removed by identifying those regions so contaminated and using a higher-bandpass median filter. 
The corresponding errors for each EW measurement are evaluated with the per-pixel standard deviation of the flux from the mean value of 
a nearby section of the continuum and propagated through the measurement.
Uncertainty in the location of the continuum is accounted for in the calculation of the per-pixel flux error from the SNR.
As an example, Figure \ref{f:rv-lc}d shows the variations in $W_\lambda$ for the Si II $\lambda$6347 $\AA$ line during the \most ~photometry. 
There is a weak correlation between the RV and EW variations during this season but it is not always present.

Comparison of ESPaDOnS/Narval spectra with 
contemporaneous AST spectra reveal that, while the instruments generally yield consistent measurements of $W_\lambda$, the 
comparatively lower-resolution 
AST spectra show a small but persistent bias in which $W_\lambda$ systematically increases with the uncertainty $\sigma(W_\lambda)$. 
However, $W_\lambda$ is accurate to within $\sim$5\% which is sufficient to reveal at times a weak correlation between RV and the $W_\lambda$ of, 
for instance, the Si {\sc II} 6347 \AA ~line, one of the stronger photospheric lines in the AST spectral window (see Fig. \ref{f:rv-lc}d). 
Figure \ref{f:residuals}  is a 50-day representative interval, occurring one year prior to the \most ~photometry, that shows the RV changes as a function of 
time (panel a).
For the same time interval, panels (c) to (e) plot the equivalent widths of five different lines.
While H$\alpha$ (having high sensitivity to wind) demonstrates the largest variation amplitude, 
the rest of the lines, though they have different absolute amplitudes, change moderately - on the order of 12 to 19\%.
The two Si \textsc{II} lines follow a similar trend and have similar amplitudes.
\section{Rigel Periodicities}\label{s:freq}
\subsection{Variability Pattern of Rigel}\label{ss:pattern}
\citet{kaufer-1996-01, kaufer-1997-01} derived several harmonics to fit to the season-by-season photometry, radial velocity and 
EW variations of Rigel and five other late BA supergiants. 
As they demonstrated, the H$\alpha$ line equivalent widths show systematic variations.
Figure 7 in \citet{kaufer-1997-01} shows that $\alpha$ Cyg, the prototype of this class of pulsating stars, and the other BSGs exhibit
variability with periods exceeding a week.
For the specific case of Rigel, the periods they find are in the range of 4 to more than 50 days.
\citet{richardson-2011-01} found evidence for season-to-season changes in the oscillation frequencies of $\alpha$ Cyg (Deneb, A2 Ia)
from five years of spectroscopic and photometric monitoring.

The need for multiple modes to fit to our RV and EW data set is clearly shown in Figure \ref{f:residuals}.
Similar behavior is observed in other seasons. 
Because different spectral lines are formed at various optical depths of the star's atmosphere, the temporal changes in $W_\lambda$
for strong lines (such as Si \textsc{II} $\lambda6347$) are significantly different from those of weak lines (such as C \textsc{II} $\lambda6578$).
The equivalent width, however, is sensitive to local changes in the effective temperature $\delta \teff/\teff$ and gravity $\delta g/g$.
In a series of studies in $\beta$ Cep and SPB stars (which are different from BA supergiants, but still have similarities in the nature of their pulsation), 
\citet{dupret-2002-01} and \citet{deridder-2002-01} showed that a sinusoidal behavior in EW and RV time series with a common frequency is observed in only few cases.
Given that Rigel (similar to SPBs) pulsates in g-modes which are transversal motions at the stellar surface, EWs are strongly affected by $\delta \teff/\teff$
\citep{deridder-2002-01}.
Based on the observed behavior of Rigel's equivalent widths (panels c to e in Figure \ref{f:residuals}), 
we believe that a complex non-radial velocity field exists across different optical depths \citep{aerts-2009-01, simon-diaz-2010-01}.
As \citet{kaufer-1997-01} showed (in their Figure 4), radial velocity variations are negligibly affected by this depth dependence.
Furthermore, the RV variations show a weak (not to say absent) degree of correlation with the EW of any of the lines.

\subsection{Frequency Analysis Based on RV data}\label{ss:freq}
\begin{figure}[t]
\epsscale{1.0} \plotone{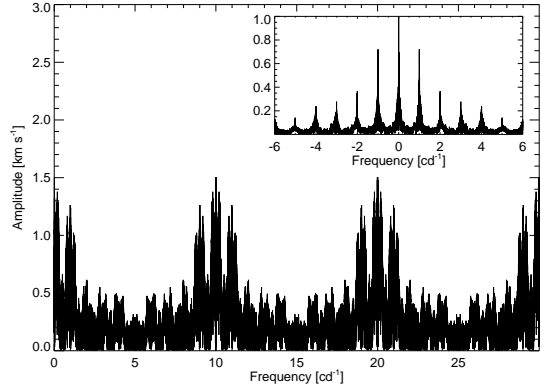}
\caption{Discrete Fourier Transform of the RV time series shows a repeating pattern at multiples of $f_{\mbox{\scriptsize nyq}}=5$ d$^{-1}$. 
The inner panel shows the spectral window and the one day aliasing in addition to the annual aliasing. \label{f:dft}}
\end{figure}

Compared to the \most ~photometry time series (Rayleigh limit of 0.036 d$^{-1}$), the RV data have a much longer time span 
(Rayleigh limit of $4.43\times10^{-4}$ d$^{-1}$). As a result, we base our search
for intrinsic signals primarily on this data set using two widely-used state-of-the-art programs - \texttt{SigSpec} 
\citep{reegen-2007-01} and \texttt{Period04} \citep{lenz-2005-01}. 
However, the RV data suffer from strong daily ($f_d=1$ d$^{-1}$) and annual 
($f_a=0.002769$ d$^{-1}$) aliasing (inner panel in Figure \ref{f:dft}). 
As shown in Figure \ref{f:dft}, there is a repeating pattern
every 5 d$^{-1}$ in the discrete Fourier transform (DFT) spectra that arises from our sampling rate. 
Thus we selected the upper frequency scan range at $f_{\mbox{\scriptsize nyq}}=$5 d$^{-1}$, and our analysis did not identify  
frequencies higher than 1 d$^{-1}$.
%
\begin{figure}[t]
\epsscale{1.0} \plotone{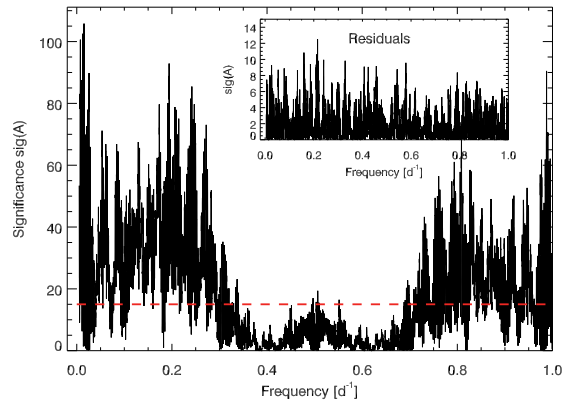}
\caption{Significance spectrum of RV data with a normalized weight. 
Only the low-frequency domain between [0, 1] d$^{-1}$ is shown.
The inner panel shows the remaining power below the threshold sig$=$15. 
After prewhitening of 19 significant harmonics, the residual is 0.94 km s$^{-1}$.
\label{f:sig-spectra}} 
\end{figure}
%
  \begin{deluxetable*}{lccccc}
  \tablecaption{Rigel periodocities found from the radial velocity study. \label{t:freq} }
  \tablecolumns{7}
  \tablehead{ \colhead{ID} & \colhead{Frequency} & \colhead{Amplitude} & \colhead{(SNR)$_{\mbox{\scriptsize RV}}$} & 
  			\colhead{sig} & \colhead{r.m.s.} \\
  			\colhead{} & \colhead{[d$^{-1}$]} & \colhead{[km s$^{-1}$]} & \colhead{} & \colhead{ } & \colhead{[km s$^{-1}$]} }
  \startdata
  $f_{1}  $& 0.01523 $\pm$ 0.00002 & 1.128 $\pm$ 0.027 & 15.9 & 105.9& 2.189 \\
  $f_{2}  $& 0.20635 $\pm$ 0.00003 & 0.964 $\pm$ 0.027 & 14.0 & 104.4& 1.946 \\
  $f_{3}  $& 0.19459 $\pm$ 0.00003 & 0.823 $\pm$ 0.027 & 11.8 & 60.0 & 1.734 \\
  $f_{4}  $& 0.02516 $\pm$ 0.00004 & 0.625 $\pm$ 0.027 & 8.9   & 46.2 & 1.617 \\
  $f_{5}  $& 0.28297 $\pm$ 0.00005 & 0.562 $\pm$ 0.027 & 8.1   & 42.8 & 1.538 \\
  $f_{6}  $& 0.06107 $\pm$ 0.00003 & 0.839 $\pm$ 0.027 & 12.0 & 36.2 & 1.465 \\
  $f_{7}  $& 0.22545 $\pm$ 0.00004 & 0.644 $\pm$ 0.027 & 9.1   & 32.1 & 1.399 \\
  $f_{8}  $& 0.01338 $\pm$ 0.00003 & 0.839 $\pm$ 0.027 & 11.9 & 27.7 & 1.341 \\
  $f_{9}  $& 0.31317 $\pm$ 0.00004 & 0.693 $\pm$ 0.027 & 10.0 & 27.9 & 1.280 \\
  $f_{10}$& 0.17352 $\pm$ 0.00005 & 0.515 $\pm$ 0.027 & 7.5 & 28.9 & 1.233 \\
  $f_{11}$& 0.09453 $\pm$ 0.00005 & 0.552 $\pm$ 0.027 & 8.0 & 20.1 & 1.180 \\
  $f_{12}$& 0.15542 $\pm$ 0.00006 & 0.452 $\pm$ 0.027 & 6.5 & 20.9 & 1.145 \\
  $f_{13}$& 0.04343 $\pm$ 0.00006 & 0.469 $\pm$ 0.027 & 6.7 & 20.4 & 1.111 \\
  $f_{14}$& 0.13921 $\pm$ 0.00004 & 0.590 $\pm$ 0.027 & 8.6 & 20.0 & 1.079 \\
  $f_{15}$& 0.12080 $\pm$ 0.00005 & 0.518 $\pm$ 0.027 & 7.5 & 22.7 & 1.046 \\
  $f_{16}$& 0.82026 $\pm$ 0.00007 & 0.404 $\pm$ 0.027 & 6.4 & 18.1 & 1.011 \\
  $f_{17}$& 0.36038 $\pm$ 0.00008 & 0.347 $\pm$ 0.027 & 5.0 & 15.3 & 0.981 \\
  $f_{18}$& 0.31699 $\pm$ 0.00007 & 0.381 $\pm$ 0.027 & 5.5 & 16.2 & 0.961 \\
  $f_{19}$& 0.51179 $\pm$ 0.00009 & 0.315 $\pm$ 0.027 & 4.7 & 16.2 & 0.938 \\
  \enddata
  \end{deluxetable*}
%

\texttt{SigSpec} computes the spectral significance level (sig) for the DFT amplitude spectrum based on the analytical solution for 
the Probability Density Function (PDF) of an amplitude level of any peak during the iterative prewhitening process. 
By default the program scans for peaks with sig$>5$ which is approximately equivalent to S/N$=3.8$ \citep{reegen-2007-01}. 
However, we conservatively chose just to search for highly significant modes to avoid a forest of low-frequency peaks. 
To accomplish this, the prewhitening process is stopped if the significance (sig)
of each peak or the cumulative significance (csig) of the whole solution is below 15. 
A similar threshold was used by \citet{chapellier-2011-01} in the frequency analysis of one of the CoRoT primary targets.
Since each RV measurement $i$ has an error $\sigma_i$, we associate a normalized 
weight $w_i=\sigma_i^{-2}/\sum_i \sigma_i^{-2}$ to each data point. 
The resulting significance spectrum is shown in Figure \ref{f:sig-spectra}. 
This procedure leads to a prewhitening of 19 significant modes. 
The corresponding list of detected harmonics is presented in Table \ref{t:freq}. 
It tabulates the multimode solution to the RV data set from \texttt{SigSpec}. 
Analytical 1-$\sigma$ uncertainties are evaluated according to \citet{montgomery-1999-01};
those in Table \ref{t:freq} are 4-$\sigma$ uncertainties.
After prewhitening, the rms of the residual is 0.94 km s$^{-1}$ and csig = 15.18.
This indicates that the probability that the presented frequency solution could be generated by noise is 1 in 10$^{\mbox{\scriptsize csig}}$.
Figure \ref{f:residuals}b plots the residuals after prewhitening. 
As shown, the residuals are small but not perfectly featureless.
To avoid misidentification of true frequencies from 
aliases, we conservatively discontinued the prewhitening at this step.
\subsection{Comparing Results of \texttt{SigSpec} and \texttt{Period04}}\label{ss:comp}
A straightforward prewhitening of the data with the same weights used as above with \texttt{Period04} starts with the same results
as in \texttt{SigSpec} for the first few harmonics, but then shows some differences. 
Instead, we imported the frequencies from the \texttt{SigSpec} list, and prewhitened sequentially. 
The maximum difference of 0.01 km s$^{-1}$ between the calculated amplitudes occurs 
in $f_{19}$. The final rms of the residual is 0.92 km s$^{-1}$ in \texttt{Perod04} and 0.94 km s$^{-1}$ in \texttt{SigSpec}. 
The signal-to-noise-ratio within a box of 1.0 d$^{-1}$ around $f_{19}$ is 4.70. 

\subsection{Flux Amplitudes}\label{ss:light}
Independent prewhitening of the \most ~dataset results in tens of modes that have no corresponding counterparts
 in the frequency list of Table \ref{t:freq}.
This is quite expected since the baseline of \most ~observations ($\sim$28 days) is shorter than some of the periods found for the star 
indicated from the analysis of the RV data.
To resolve this, we used the ``fixed" frequencies derived from the RV analysis (Table \ref{t:freq})  and employed \texttt{Period04} to determine 
the possible corresponding light amplitudes in the \most ~data.
We searched for stable amplitudes and phases (an additional harmonic at the orbital 
frequency of the satellite $f_{\mbox{\scriptsize orb}}=14.19827$ d$^{-1}$ was also subtracted). 
Unfortunately, due to the presence of long-period modes comparable to (and longer than) the \most ~photometry baseline, 
no converging solution could be achieved.
\begin{figure*}[t]
\epsscale{1.0} \plotone{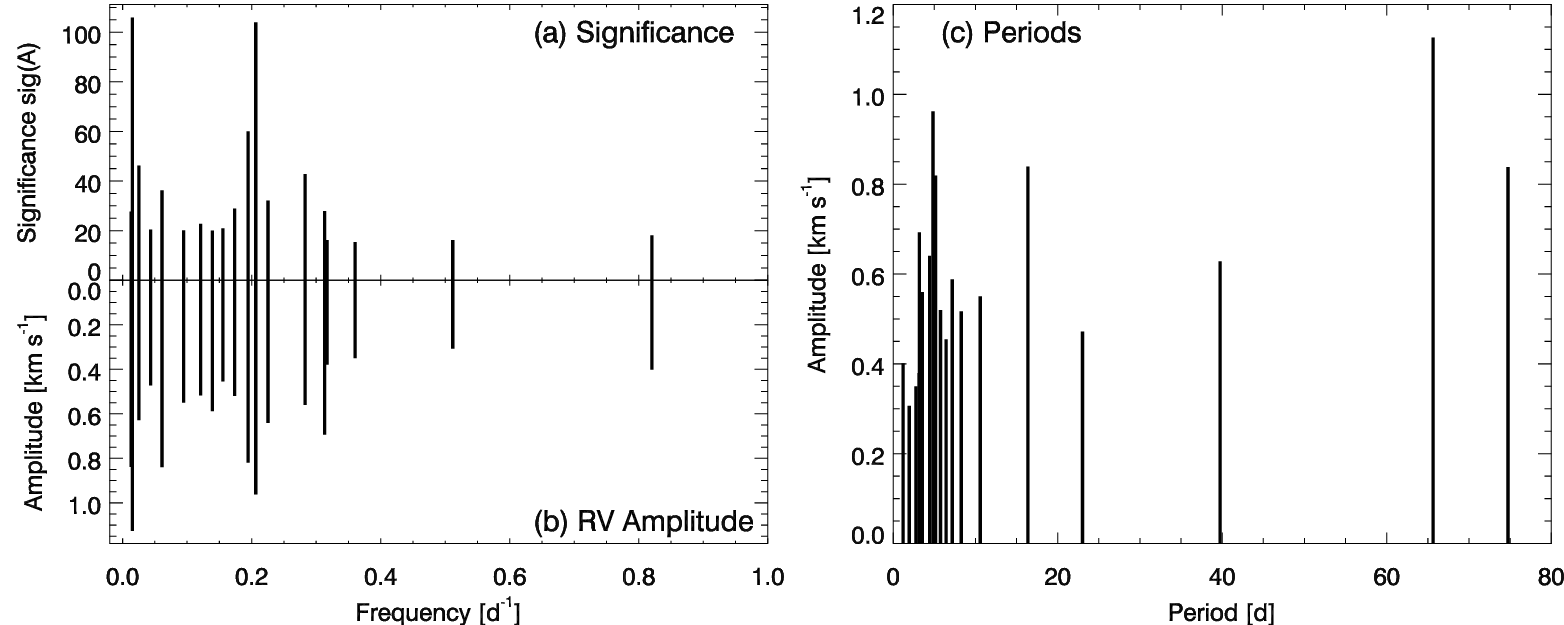}
\caption{Left. Detected frequencies with their corresponding significance level (panel a) and radial velocity amplitude (panel b).
The multimode solution is tabulated in Table \ref{t:freq}.
Right. Period distribution of detected modes with their corresponding amplitude.
\label{f:result}} 
\end{figure*}
%
\subsection{Frequency Distribution of Multimode Solution}\label{ss:dist}
According to the asymptotic theory of non-radial stellar oscillations \citep{tassoul-1980-01}, high order p- and g-modes present
regularity in frequencies and periods, respectively \citep{aerts-2010-01}.
In the case of Rigel and similar classes of stars, which are believed to be unstable against g-modes, this regularity in period can 
bring a wealth of information about how stellar material is mixing in the stars' deep interior \citep[see][for an example]{degroote-2010-01}.
However, as a prerequisite for applying this technique, identification of polar and azimuthal degrees ($\ell, m$) for each individual mode is necessary.
Unfortunately, the number of secured frequencies is not large enough for a statistical study of any possible period spacings.
Moreover, the mode identification is beyond the scope of this paper.

However, after extraction of the frequencies with good precision, it is worthwhile to demonstrate how the modes are distributed according to their
corresponding frequencies and amplitudes.
For this purpose, the left panel in Figure \ref{f:result} shows the distribution of detected frequencies between [0, 1] d$^{-1}$ and 
their corresponding RV amplitudes. 
Except $f_{16}$ and $f_{19}$, the rest of the modes have frequencies below 0.4 d$^{-1}$. 
The right panel of Figure \ref{f:result} shows the corresponding period distribution.

It is possible that the rotation period of Rigel could be detected in the frequency analyses of the RV data. 
\citet{papics-2011-01} found the rotational modulation as the only explanation for the variability in the B0.5 IV CoRoT double-line spectroscopic binary
HD 51756.
Rigel has two modern, self-consistent spectroscopic $v \sin i$ measures \citep{przybilla-2010-01, simon-diaz-2010-01} 
from which we adopt $v \sin i$ = 25 km s$^{-1}$. 
This $v \sin i$ is typical for a blue supergiant  and indicates that an inclination of $i> 60^\circ$ is likely 
(the inclination $i$ is defined in the usual way as the angle of star's rotation axis relative to our line of sight). 
Adopting both $i = 60^\circ$ and $i = 90^\circ$ with $R = 78.9 R_\odot$, yield
$P_{\mbox{\scriptsize rot}}(i=60^\circ) = 137$ d ($\equiv$0.0073 d$^{-1}$) and $P_{\mbox{\scriptsize rot}}(i=90^\circ) = 158$ d 
($\equiv$0.0063 d$^{-1}$), respectively.  
The value of  $i >60^\circ$ is also in accord with the simple angular momentum conservation assumption from the 
expected $v_{\mbox{\scriptsize rot}} $ = 240-300 km s$^{-1}$ estimated for Rigel when it was a main-sequence O9/B0 V star.  
As shown in Table \ref{t:freq}, there are no significant frequencies in the range of 0.006 - 0.007 d$^{-1}$. 
So, rotation effects can be neglected in our interpretation of frequencies.
Yet, we are aware that multiple integers of the rotational frequency might show up in the frequency analysis, and long-period modes 
could have a non-pulsational origin.
Given the weak magnetic field on Rigel, rotational modulations induced by stellar spots are unlikely.

\subsection{Frequency Analysis of Equivalent Widths}
As panels c to e in Figure \ref{f:residuals} clearly show, Si \textsc{II} $\lambda6347$, Si \textsc{II} $\lambda6371$, He \textsc{I} 6678 and H$\alpha$
exhibit the large amplitude changes in EW. 
However, the binned EW measures have a very low duty cycle and DFT analyses do not lead to any significant frequencies above the 
4-$\sigma$ noise threshold.
There is only one mode detected for the Si \textsc{II} $\lambda6347$ line with frequency, amplitude and SNR of
0.0336 d$^{-1}$, 14.71 m\AA ~and 4.24, respectively.
Similarly for Si \textsc{II} $\lambda6371$ line, the only dominant peak in the Fourier power spectrum corresponds to the frequency, 
amplitude and SNR of 0.0336 d$^{-1}$, 11.32 m\AA ~and 3.46, respectively.
Therefore, not only do the frequencies from EWs show no apparent correspondences with any of the frequency entries in Table \ref{t:freq}, 
the first two smallest frequencies, i.e. $f_8$ and $f_1$, do not appear in the DFT power spectrum of EWs.
Hence, with the current time series of spectra at our disposal, we cannot associate the low frequency range of RV variations with EW variations.
%
\section{Summary}
This is the first time that simultaneous space-based photometry and time-resolved medium-resolution optical spectroscopy of a blue supergiant star
are presented.
Although the short baseline of \most ~photometry did not enable us to derive flux amplitudes, the long baseline of RV monitoring and the DFT
analysis revealed 19 significant modes above SNR=4.6.
These have been shown to arise from non-radial pulsations by \citet{kaufer-1997-01} with gravity-mode nature \citep{ lefever-2007-01}.

The question of the degree of regularity in light and RV variability in BA supergiants remains unanswered until an uninterrupted 
long time-baseline space-based observation campaign of these objects is carried out. 
It is possible that a degree of semi-regularity might exist in the current RV data set.
But, there is no definite evidence for this, and thus it was not included in our frequency extraction procedure.
Therefore, we caution that the periodocities found in our analysis might be affected by this issue.
In the near future it will be possible for long term photometry of Rigel and other bright BSGs to be carried out over several months by the 
BRITE-Constellation Mission \citep{kuschnig-2009-01}. 
Observations like these should result in better defined frequencies.

The temporal modulations in the equivalent widths of the two Si \textsc{II} lines show a completely different pattern from those of He \textsc{I} 
$\lambda6678$ and H$\alpha$ line, which indicate that equivalent widths formed at different optical depths $\tau_\lambda$ are uncorrelated.
This demonstrates that there exists a pulsationally induced non-radial velocity gradient across different optical depths from which the lines 
are formed \citep{aerts-2009-01, simon-diaz-2010-01}.
As a result, the response of the dilute extended atmosphere of Rigel (which is reminiscent of other BA supergiants) to the pulsational waves
emerging from the interior of the star is very complex, and has a high potential to probe a wide range of depths where the spectral lines are formed.
As shown in Figure \ref{f:residuals}d, the equivalent width of the H$\alpha$ feature shows the largest systematic variations of any of the spectral features
measured.
Because the H$\alpha$ feature is also affected by mass outflows/inflows, and winds, it is a difficult line to interpret.
It is, therefore, not surprising that its behavior is different than the other spectral lines studied.
A full asteroseismic study of the Rigel is planned in a subsequent paper.

\acknowledgments 
\textbf{Acknowledgments} We are grateful to the referee, Nancy Morrison, for her valuable comments in improving this paper. 
EM appreciates the hospitality and support he received from Wojcieh Dziembowski during his visit to Nicolas Copernicus 
Astronomical Center in Warsaw where a part of this study was done; he is also grateful to Patrick Lenz and Eric Chapellier for fruitful discussions on 
frequency analysis of an uneven sample. 
We thank Joel Eaton and Frank Fekel for coordinating the AST spectroscopic observations.
The research at Tennessee State University was supported in part by NASA, NSF, Tennessee State
University, and the state of Tennessee through its Centers of Excellence program.
We also like to thank Jaymie Matthews and the \most ~team for acquiring and processing the photometry.
This work is supported by NASA/MOST grant NNX09AH28G.

\textbf{Facilities:} \facility{TSU:AST, \most}
\bibliographystyle{apj}
\bibliography{apj-jour,../../my-bib}
\end{document}